\documentclass[%
 reprint,
 amsmath,
 aps, prb,
]{revtex4-2}

\usepackage[protrusion=false]{microtype} % Improves kerning of letter forms
\usepackage[dvipsnames]{xcolor}
\usepackage{xspace}
\usepackage[normalem]{ulem}

\usepackage{graphicx}
\graphicspath{{figures/}}
\usepackage{dcolumn} % Align table columns on decimal point
\usepackage{physics}
\usepackage{mathtools}
\usepackage[version=3]{mhchem}

\setcitestyle{super}
\usepackage[colorlinks=true,linkcolor=blue,breaklinks]{hyperref}
\usepackage{cleveref}
\crefname{figure}{Figure}{Figures}
\crefname{subfigure}{Figure}{Figures}
\crefname{table}{Table}{Tables}
\crefname{equation}{Eq.}{Eqs.}
\creflabelformat{equation}{#2\textup{#1}#3}
\crefname{section}{Section}{Sections}
\crefname{subsection}{Section}{Sections}
\crefname{subsubsection}{Section}{Sections}

\usepackage[normalem]{ulem}

\newcommand{\subt}[1]{\textsubscript{#1}}

\newcommand{\expn}[1]{\ {\operatorname{exp}}\left[{#1}\right]}

\newcommand{\bR}{\boldsymbol{R}}

\newcommand{\hH}{\hat{H}}
\newcommand{\hU}{\hat{U}}
\newcommand{\hh}{\hat{h}}
\newcommand{\hr}{\hat{r}}

\newcommand{\hmH}{\hat{\mathcal{H}}}
\newcommand{\hmX}{\hat{\mathcal{X}}}

\newcommand{\hbmu}{\mkern2mu\hat{\mkern-2mu\boldsymbol{\mu}}}
\newcommand{\hbmX}{\mkern4mu\hat{\mkern-4mu\boldsymbol{\mathcal{X}}}}

\usepackage{xcolor}
\usepackage{physics}
\usepackage[outline]{contour} % glow around text

\usepackage{tikz}
\usepackage{circuitikz}
\usepackage{quantikz}
\usepackage{pgfgantt}

\usepackage{siunitx}

\usepackage{pgfplots}
\pgfplotsset{compat=newest}

\usetikzlibrary{3d}
\usetikzlibrary{angles,quotes} % for pic (angle labels
\usetikzlibrary{patterns} % Include this line to enable pattern lib)
\usetikzlibrary{snakes}
\usetikzlibrary{calc, decorations.pathreplacing, decorations.markings} %fadings,
\tikzset{>=latex} % for LaTeX arrowhead

\usetikzlibrary{fit, positioning}
\usetikzlibrary{intersections, patterns, pgfplots.fillbetween}
\usetikzlibrary{arrows, arrows.meta, shapes.geometric}

\usetikzlibrary{hobby} % for ..

\definecolor{tensorblue}{rgb}{0.8,0.8,1}
\definecolor{tensorred}{rgb}{1,0.5,0.5}
\definecolor{tensorpurp}{rgb}{1,0.5,1}

\tikzset{ten/.style={fill=tensorblue}}
\tikzset{tenred/.style={fill=tensorred}}
\tikzset{tengreen/.style={fill=green!50!black!50}}
\tikzset{tenpurp/.style={fill=tensorpurp}}
\tikzset{tengrey/.style={fill=black!20}}
\tikzset{tenorange/.style={fill=orange!30}}
\tikzset{u/.style={fill=blue!20,draw=black}}
\tikzset{w/.style={fill=green!50!black!80,draw=black}}

\pgfdeclarelayer{l1}
\pgfdeclarelayer{l2}
\pgfsetlayers{l1,l2,main}

\definecolor{myred}{HTML}{FF1F5B}
\definecolor{myblue}{HTML}{009ADE}
\definecolor{mygreen}{HTML}{00CD6C}
\definecolor{myorange}{HTML}{F27522}
\colorlet{mucol}{red!90!black}
\colorlet{agg_col}{blue!80!}

\tikzstyle arrowstyle=[scale=1]
\tikzstyle directed=[postaction={decorate,decoration={markings,
    mark=at position .65 with {\arrow[arrowstyle]{stealth}}}}]
\tikzstyle reverse directed=[postaction={decorate,decoration={markings,
    mark=at position .65 with {\arrowreversed[arrowstyle]{stealth};}}}]

\tikzset{>=latex} % for LaTeX arrow head

\tikzstyle{vector_red}=[->,line width=2.0pt,mucol]
\tikzstyle{vector_blue}=[->,line width=2.0pt,myblue]
\tikzstyle{dashed_red}=[->,line width=2.0pt, mucol, dashed]

\makeatletter
\pgfkeys{%
  /tikz/node on layer/.code={
    \gdef\node@@on@layer{%
      \setbox\tikz@tempbox=\hbox\bgroup\pgfonlayer{#1}\unhbox\tikz@tempbox\endpgfonlayer\egroup}
    \aftergroup\node@on@layer
  },
  /tikz/end node on layer/.code={
    \endpgfonlayer\endgroup\endgroup
  }
}

\def\node@on@layer{\aftergroup\node@@on@layer}
\makeatother

\tikzset{%
  >={Latex[width=2mm,length=2mm]},
            base/.style = {rectangle, rounded corners, draw=black,
                           minimum width=4cm, minimum height=1cm,
                           text centered}, %font=\rmfamil},
  activityStarts/.style = {base, fill=blue!20},
       startstop/.style = {base, fill=red!30},
    activityRuns/.style = {base, fill=green!30, minimum size=2mm},
         process/.style = {draw, fill=orange!15, dashed, inner sep = 5pt, %minimum width=0.5cm, 
                           },%font=\ttfamily},
  lvl1/.style={draw,fill=red!20,rounded corners=0.2cm,inner sep=5pt,node on layer=l1,  minimum size=2mm},
  lvl2/.style={draw,fill=blue!25,rounded corners=0.2cm,inner sep=5pt,node on layer=l2,  minimum size=2mm},
  lvl3/.style={draw=blue,fill=white,dashed,rounded corners=0.25cm,align=flush center,text width=12em,inner sep=4pt,minimum height=1.5cm}, 
  title/.style={node font=\large,  minimum size=1mm},
  myarrow/.style={latex-latex,ultra thick,blue!80},
}

\tikzset{circle with arrow/.style args={#1,#2}{
        circle, draw=none, 
        thick, inner sep=1pt,
        minimum size=12pt, fill=#1,
        font=\sffamily\bfseries\color{#2}
    },
    arrow style/.style={
        ->, line width=1.4pt, % make the arrow thicker
        >=latex,
        white
    }
}

\begin{document}

\preprint{}

\title{Generalized Quantum Electrodynamics Self-Consistent Field Theory For Strong Light-Matter Interactions}

\author{Ilia M. Mazin}
\affiliation{%
  Theoretical Division, Los Alamos National Laboratory, Los Alamos, New Mexico 87545, United States
}%
\author{Yu Zhang}
  \email{zhy@lanl.gov}
\affiliation{%
  Theoretical Division, Los Alamos National Laboratory, Los Alamos, New Mexico 87545, United States
}%

\date{\today}

\begin{abstract}
\textit{Ab initio} theory is essential for providing atomic-level insight into polariton-mediated emergent phenomena, in which strong light--matter coupling gives rise to hybrid quantum states with both electronic and photonic character. In this work, we develop a generalized quantum electrodynamics self-consistent field (QED-SCF) theory that extends prior QED Hartree--Fock (QED-HF) frameworks to include an arbitrary number of photonic Fock states, moving beyond the conventional restriction to the photon vacuum. Our formulation unifies and generalizes existing methods---including QED-HF, strongly coupled QED-HF (SC-QED-HF), and variationally transformed QED-HF (VT-QED-HF). Through systematic numerical comparisons, we demonstrate that SC-QED-HF and VT-QED-HF incorporate electron--photon correlation while requiring significantly fewer Fock states to achieve convergence, owing to their variationally entangled ans{\"a}tze. These results underscore the importance of physically motivated photonic transformations for improving the efficiency and accuracy of \textit{ab initio} QED approaches and offer a scalable route to first-principles simulations of strongly coupled light--matter systems.
\end{abstract}

\maketitle

\section{Introduction}
\label{sec:intro}

Light--matter interactions are fundamental to a wide range of physical and chemical phenomena, from natural processes such as photosynthesis to applications in optical communication. These interactions can be broadly categorized according to their strength. In the weak-coupling regime, light and matter largely retain their individual properties. In the strong-coupling regime, however, hybrid quasiparticles with mixed light--matter character, known as ``polaritons", form~\cite{Rider2021:p217,Rivera2020:p1368,Herrera2020}.
Experiments in polaritonic chemistry~\cite{Ebbesen2016:p2403,Kockum2019:p19,Rivera2020:p1368,Herrera2020,Simpkins2023,Xiang2024:p2512,Hsu2025:p1604} have demonstrated control over chemical processes and photophysics through strong light--matter coupling. The ability to manipulate electronic structure, chemical reaction rates, and alter photochemical pathways~\cite{Hutchison2012:p1592,Ribeiro2018:p6325,Mony2021:p2010737,Zeng2023:p19655,Lee2024:p9544,Thomas2024:p2309393,Koessler2025:p11644,Mandal2019JPCL,Zhang2019JCP,Fregoni2018NC} without changing the underlying chemical composition has potential applications in catalysis,~\cite{MartinezMartinez2017AP,Galego2015PRX} materials design,~\cite{Latini2019NL, Schlawin2022, zhang2026chiral} and quantum technologies.~\cite{Kavokin:2022tn}

Fully realizing the potential of polaritonic chemistry requires understanding the interactions between molecular systems and electromagnetic fields in coupled environments. \textit{Ab initio} methods~\cite{Szabo1982,Helgaker2000,Fetter2003,Dickhoff2008,Mandal:2023vh,Weight2023pccp} help elucidate the microscopic processes driving these phenomena and bridge theory and experiment. Perturbation theory~\cite{Rayleigh:1894,Shavitt:2009} provides well-established methods for treating weak light--matter coupling as a small perturbation to the electronic structure. In the strong-coupling regime, however, significant state hybridization necessitates a nonperturbative treatment of light and matter on an equal footing.~\cite{Schaefer2018:p043801,Ashida2021PRL,Bernardis2018PRA}

The number of \textit{ab initio} methods designed to describe cavity QED phenomena has grown steadily over the past few decades.~\cite{Flick2017:p3026,Schaefer2018:p043801,foley:2023CPR,Mandal:2023vh,Flick2020JCP,Rivera:2019tr,Bauman2025:p10035,Weight2024:p032804} QED Hartree--Fock~\cite{Haugland2020:p041043,Bauman2025:p10035,Riso2022:p0,Li2023:p0,Mazin:2024p0} (QED-HF) and quantum electrodynamical density functional theory~\cite{Ruggenthaler2014:p012508,Tokatly2013PRL,yang_quantum-electrodynamical_2021,Yang:2022va} (QEDFT) extend their traditional electronic-structure counterparts by incorporating the quantized electromagnetic field into the molecular Hamiltonian, thereby providing a mean-field description of the electron--photon interaction. Methods such as strongly coupled QED-HF~\cite{Riso2022:p0,Moutaoukal:2024p8911,Castagnola2025:p4447} (SC-QED-HF), variationally transformed QED-HF~\cite{Li2023:p0} (VT-QED-HF), and variationally squeezed QED-HF~\cite{Mazin:2024p0} (VSQ-QED-HF) incorporate correlation effects beyond the product-state mean-field description by introducing additional variational parameters associated with the photonic degrees of freedom. Related variational and polaron-transformation approaches provide additional context for these photonic transformations.~\cite{Rivera:2019tr,Silbey1984jcp,lang1963kinetic,Luo_polaron:2022prb,Cui:2024p1143}
Post-HF methods, such as QED coupled-cluster~\cite{Haugland2020:p041043,deprince_cavity-modulated_2021} (QED-CC) and QED configuration interaction~\cite{McTague2022:p154103} (QED-CI), provide more accurate treatments of electron--photon correlation at greater computational cost. Other correlated and response-based approaches have likewise been formulated for cavity QED systems.~\cite{flick_lightmatter_2019,Flick2020JCP,Weight2023pccp,Castagnola2025:p4447,DePrince2025:p8024} Despite these advances, many practical implementations restrict the photonic component of the wavefunction to the vacuum state. Methods that accommodate variable photon populations are therefore needed.

In this work, we develop a generalized QED self-consistent-field (QED-SCF) framework for strongly coupled light--matter systems that extends QED Hartree--Fock theory beyond the conventional photon-vacuum approximation by allowing an arbitrary number of photonic Fock states. The formulation unifies QED-HF, SC-QED-HF, and VT-QED-HF within a common theoretical and computational framework, which we implement in \textsc{OpenMS}, our open-source Python package. We demonstrate and assess the framework using two molecular systems across a broad range of light--matter coupling strengths, with particular attention to systematic convergence with respect to the photonic basis size. The results show how physically motivated photonic transformations incorporate electron--photon correlation and substantially reduce the number of Fock states needed to converge the ground-state energy. Although this study focuses on cavity QED, \textsc{OpenMS} is under active development to encompass a broader range of light--matter interactions and quantum-dynamics phenomena. The remainder of the paper presents the theoretical background and working equations, describes the computational details, discusses the numerical results, and summarizes their implications for simulations of complex light--matter interactions.

\section{\label{sec:theory}Theory}
In this section, we introduce the principal components of QED Hartree--Fock methods and define the notation used throughout this work. We consider a molecular system interacting with a single cavity mode. The total number of electrons is denoted by ${N_{\text{e}}}$, and each atom $\text{A}$ has nuclear charge ${Z}_{\text{A}}$. The operators $\hat{a}^\dagger_{p}$ and $\hat{a}_{p}$ create and annihilate electrons, respectively, whereas $\hat{b}^\dagger_{\alpha}$ and $\hat{b}_{\alpha}$ create and annihilate photons in mode $\alpha$.

\subsection{Pauli-Fierz Hamiltonian}
The Pauli-Fierz (PF) Hamiltonian~\cite{Pauli1937:p572,Power1959PTRSA,Woolley2020} is routinely used to describe light-matter interactions in molecular quantum electrodynamics. In this work, we employ the PF Hamiltonian formulated in the length gauge, derived under the long-wavelength approximation~\cite{Rokaj2018:p034005,Schaefer2020:p975}, which assumes the wavelength of electromagnetic radiation to be much larger than the spatial extent of the molecular system.
\begin{align}
    \label{eq:pf_h}
    {\hH}^{}_{\text{PF}}
        = {\hH}^{}_{\text{e}} + {\hH}_{\text{ph}}
          + {\hH}_{\text{DSE}} + {\hH}_{\text{e-ph}}
    \subt{.}
\end{align}
The first term of Eq.~\ref{eq:pf_h} is the standard electronic Hamiltonian within the Born--Oppenheimer approximation,
\begin{align}
    \label{eq:elec_h}
    {\hH}^{}_{\text{e}}
        = \sum_{pq} h^{}_{pq} \hat{a}^\dagger_{p} \hat{a}_{q}
          + \frac{1}{2} \sum_{pqrs} v^{}_{pqrs} \hat{a}^\dagger_{p} \hat{a}^\dagger_{q} \hat{a}_{s} \hat{a}_{r}
    \subt{,}
\end{align}
expressed in terms of the standard one-electron integrals (OEI)
$h_{pq}= \int {\psi}^{*}_{p} (1) {\hh}_{1} {\psi}^{}_{q} (1) \dd{{\tau}_{1}}$ and electron repulsion integrals (ERI), 
$v_{pqrs}= \int {\psi}^{*}_{p} (1) {\psi}^{}_{q} (1) 
\hr^{-1}_{\text{12}} 
{\psi}^{*}_{r} (2) {\psi}^{}_{s} (2)
\dd{{\tau}_{1}} \dd{{\tau}_{2}}$, respectively.
The remaining terms of Eq.~\ref{eq:pf_h} are the photonic ${\hH}_{\text{ph}}$, dipole self-energy ${\hH}_{\text{DSE}}$, and electron-photon ${\hH}_{\text{e-ph}}$ interaction components, which depend on properties of the cavity mode, including the frequency ${\omega}_{\alpha}$, molecule-cavity coupling strength ${\lambda}_{\alpha}$, and polarization vector ${\boldsymbol{e}}_{\alpha}$,
\begin{align}
    \label{eq:pf_p_terms}
    {\hH}_{\text{ph}}
        &= \sum_\alpha {\omega}_{\alpha} \hat{b}^\dagger_{\alpha} \hat{b}_{\alpha}
    \subt{,} \\
    \label{eq:pf_dse_terms}
    {\hH}_{\text{DSE}}
        &= \frac{1}{2} \sum_\alpha  (\boldsymbol{\lambda}_\alpha \cdot \hat{\boldsymbol{\mu}})^{2}
    \subt{,} \\
    \label{eq:pf_ep_terms}
    {\hH}_{\text{e-ph}}
        &= \sum_\alpha \sqrt{\frac{\omega_\alpha}{2}}
           (\boldsymbol{\lambda}_\alpha \cdot \hat{\boldsymbol{\mu}})
           ( \hat{b}^\dagger_{\alpha} + \hat{b}_{\alpha} )
    \subt{,}
\end{align}
where $\boldsymbol{\lambda}_{\alpha} \equiv {\lambda_{\alpha}} \boldsymbol{e}_{\alpha}$ is the mode-coupling vector, defined as the normalized mode-polarization vector scaled by the coupling strength.
The DSE and electron--photon terms in Eqs.~\ref{eq:pf_dse_terms} and \ref{eq:pf_ep_terms} are written in terms of the molecular dipole operator, ${\hat{\boldsymbol{\mu}}} =\sum_{pq} \left(\delta^{}_{pq} \frac{\boldsymbol{\mu}^{\text{nuc}}}{N_{\text{e}}}
- \boldsymbol{\mu}^{\text{e}}_{pq}\right) \hat{a}^\dagger_{p} \hat{a}_{q}$,
defined with respect to the nuclear and electronic degrees of freedom, $\boldsymbol{\mu}^{\text{nuc}}
        \equiv \sum\limits_{\text{A}}^{\mathclap{N_{\text{A}}}}
                {Z}_{\text{A}} {\bR}_{\text{A}}$ and $\boldsymbol{\mu}^{\text{e}}_{pq}
        \equiv \mel{{\psi}_{p}}{\boldsymbol{r}}{{\psi}_{q}}$, respectively.

The ${\hH}_{\text{DSE}}$ component of the PF Hamiltonian represents the self-interaction of the molecular dipole moment with the quantized electromagnetic field. Its inclusion is essential for a stable and gauge-consistent description of the light--matter Hamiltonian, particularly beyond the weak-coupling regime~\cite{Rokaj2018:p034005,Schaefer2020:p975,Stefano2019NP,Garziano2020PRA}. To rewrite the DSE Hamiltonian in a format similar to the electronic component, one can expand the square of the dipole operator ${\hat{\boldsymbol{\mu}}}$ as
\begin{align}
    \label{eq:dse_square_complete}
    {\hH}_{\text{DSE}} = \frac{1}{2}
    \Bigl(
        \sum_{pqrs} g^{\alpha}_{pq} g^{\alpha}_{rs}
        \hat{a}^\dagger_{p} \hat{a}^\dagger_{q} \hat{a}_{s} \hat{a}_{r}
        - \sum_{pq} Q^{\alpha}_{pq} \hat{a}^\dagger_{p} \hat{a}_{q}
    \Bigr),
\end{align}
written in terms of the dipole-coupling matrix $g^{\alpha}\equiv ( \boldsymbol{\lambda}_{\alpha} \cdot \boldsymbol{\mu}^{}_{pq} )$ and a matrix of modified quadrupole integrals $Q^{\alpha}\equiv ( \boldsymbol{\lambda}_{\alpha} \cdot \boldsymbol{Q}^{}_{pq} \cdot \boldsymbol{\lambda}_{\alpha})$. This expression makes no assumptions about the completeness of the basis set~\cite{foley:2023CPR}.
Alternatively, assuming a complete basis set, the DSE Hamiltonian can be written as
\begin{equation}
    \hH_{\text{DSE}} = \frac{1}{2}
    \left(\sum_{pqrs} g^{\alpha}_{pq} g^{\alpha}_{rs} \hat{a}^\dagger_{p} \hat{a}^\dagger_{q} \hat{a}_{s} \hat{a}_{r}
        - \sum_{pqr} g^{\alpha}_{pr} g^{\alpha}_{rq}\hat{a}^\dagger_{p} \hat{a}_{q} \right).
\end{equation}
Nevertheless, the DSE Hamiltonian can be readily absorbed into the electronic Hamiltonian with modified OEI and ERI in practical implementations.

The electron--photon component ${\hH}_{\text{e-ph}}$ is the bilinear light--matter interaction in the length-gauge Pauli--Fierz representation and can be expressed in terms of the dipole integrals as~\cite{Schaefer2018:p043801}
\begin{align}
    \label{eq:eph_ham}
    {\hH}_{\text{e-ph}}
        &= \sum_{\alpha} \sqrt{\frac{\omega_{\alpha}}{2}}
            \cdot \sum_{pq} g^{\alpha}_{pq} \hat{a}^\dagger_{p} \hat{a}_{q}
            ( \hat{b}^\dagger_{\alpha} + \hat{b}_{\alpha} ).
\end{align}
This term involves the photonic operator $(\hat{b}^\dagger_{\alpha} + \hat{b}_{\alpha})$ and therefore contributes variationally only when the photonic basis includes states beyond the photon vacuum (i.e., photon occupation number $n>0$).

\subsection{Wavefunction Ans{\"a}tze and Unitary Transformations}
A generalized many-body ansatz for the QED wavefunction of a molecular-cavity system is defined by applying a unitary operator to a product state of electronic and photonic subsystems,
\begin{align}
    \label{eq:qed_wf}
    \ket{\Psi}
        \equiv \hU \ket{{\Psi}^{}_{\text{e}}}
        \otimes \ket{{\Psi}^{}_{\text{ph}}}
    \subt{,}
\end{align}
where $\hU$ can be a functional of both subsystems, introducing entanglement between the electronic and photonic degrees of freedom. The electronic reference state $\ket{{\Psi}^{}_{\text{e}}}$ can be any single- or multi-determinant electronic many-body wavefunction. The photonic wavefunction $\ket{{\Psi}_{\text{ph}}}$ for the quantized electromagnetic fields of cavity modes $\ket{\alpha}$ is expressed as
\begin{align}
    \label{eq:p_wf}
    \ket{{\Psi}^{}_{\text{ph}}}
        = \prod_{\alpha} \sum^{n_{\alpha}}_{k=0}
        \mathcal{C}^{}_{\alpha k}
        \big( \hat{b}^\dagger_{\alpha} \big)^{k}
        \ket*{{0}^{\alpha}_{\text{ph}}},
\end{align}
a linear combination of the mode vacuum state $\ket*{{0}^{\alpha}_{\text{ph}}}$ and Fock states up to the truncation ${n}_{\alpha}$. For ${n}_{\alpha}=0$, the mode wavefunction reduces to the vacuum state $\ket*{{0}^{\alpha}_{\text{ph}}}$.

{\bf Fock State Representation.} If $\hU$ is the identity, Eq.~\ref{eq:qed_wf} is simply a direct product of electronic and photonic states, with a Fock state representation for the photonic degrees of freedom. We will later demonstrate that this ansatz requires a large number of Fock states to achieve energy convergence. This issue is especially pronounced for charged molecules, where the Fock state representation induces an origin-dependence error under finite truncation~\cite{foley:2023CPR,Matousek2024:p9424}. Moreover, as the molecule is displaced from the origin, the computational cost increases significantly, rendering this representation impractical for modeling collective light–matter coupling regimes.

{\bf Coherent State Representation.} In the special case where the photonic reference state is the vacuum $\ket*{{\Psi}^{}_{\text{ph}}}\equiv\ket*{{0}^{\alpha}_{\text{ph}}}$ and $\hU$ is chosen to be the displacement operator $\hat{D} (z_{\alpha})$,
\begin{align}
    \label{eq:cs_op}
    \hU \equiv \hat{D} ( z_{\alpha} )
        &= \expn{z_{\alpha} ( \hat{b}^\dagger_{\alpha} - \hat{b}_{\alpha} )},
\end{align}
With the displacement $z_{\alpha} = -\frac{\boldsymbol{\lambda}_{\alpha} \cdot \expval*{\hat{\boldsymbol{\mu}}}}{\sqrt{2 \omega_\alpha}}$, the wavefunction in Eq.~\ref{eq:qed_wf} is transformed into the coherent-state representation.~\cite{Philbin2014:p742} The parameter $z_\alpha$ defines the coherent shift of cavity mode $\alpha$ due to its coupling to the molecular subsystem and is proportional to the expectation value of the molecular dipole operator, $\expval*{\hat{\boldsymbol{\mu}}}$. The coherent-state representation removes explicit electron--photon interaction terms from the Pauli--Fierz Hamiltonian by absorbing the coupling into the displaced photonic vacuum. This transformation does not, however, entangle the electronic and photonic subsystems.

{\bf Entangled Representations.} 
A more general form of the unitary operator $\hU$ is defined by the displacement operator $\hat{D} ( \hat{\boldsymbol{\mu}}, f_{\alpha} )$:
\begin{align}
    \label{eq:gen_unitary}
    \hU \equiv \hat{D} ( \hat{\boldsymbol{\mu}}, f_{\alpha} )
        &= \expn{ \frac{-f_{\alpha} ( \boldsymbol{\lambda}_{\alpha} \cdot \hat{\boldsymbol{\mu}} )}
                        {\sqrt{2 {\omega}_{\alpha}}}
                ( \hat{b}^\dagger_{\alpha} - \hat{b}_{\alpha} )
            }
    \subt{.}
\end{align}
The unitary transformation in Eq.~\ref{eq:gen_unitary} depends explicitly on the full molecular dipole operator $\hat{\boldsymbol{\mu}}$ and on a variational parameter $f_{\alpha} \in [0,1]$ that controls the extent of the transformation. This parameter allows the ansatz to interpolate between different cavity-coupling regimes and yields an optimized mean-field solution. The $f_{\alpha}$ parameters were introduced in variationally transformed QED Hartree--Fock~\cite{Li2023:p0} (VT-QED-HF), and a similar variational transformation was proposed independently by Cui and co-workers.~\cite{Cui:2024p1143} In the limit $f_{\alpha}=1$, the ansatz reduces to the strongly coupled QED Hartree--Fock (SC-QED-HF) formulation developed by Riso and co-workers.~\cite{Riso2022:p0,Moutaoukal:2024p8911} If $f_{\alpha}=0$, the original unentangled product-state ansatz is recovered.

The unitary transformation $\hU$ acts on the Pauli--Fierz Hamiltonian to yield the transformed Hamiltonian $\hmH_{\text{PF}} \equiv \hU^{\dagger} \hat{H}_{\text{PF}} \hU$,
\begin{align}
    \label{eq:eff_pf_ham}
    \hmH_{\text{PF}}
        = & \hH^{}_{\text{ph}}
        + \hat{\mathcal{H}}_{\text{e}} \big[ \hbmX \big]
        + \frac{1}{2} (\Delta \boldsymbol{\lambda}_{\alpha} \cdot \hat{\boldsymbol{\mu}} {)}^{2}
    \nonumber \\
        & + \sqrt{\frac{{\omega}_{\alpha}}{2}}
            ( \Delta\boldsymbol{\lambda}_{\alpha} \cdot \hat{\boldsymbol{\mu}} ) ( \hat{b}^\dagger_{\alpha} + \hat{b}_{\alpha}),
\end{align}
where $\Delta\boldsymbol{\lambda}_{\alpha} = (1-f_\alpha)\boldsymbol{\lambda}_{\alpha}$, $\hH_{\text{ph}}$ is the original photonic component of the Pauli--Fierz Hamiltonian [Eq.~\ref{eq:pf_p_terms}], and $\hmH_{\text{e}}\big[ \hbmX \big] \equiv \hU^{\dagger} \hat{H}_{\text{e}} \hU$ is the electronic Hamiltonian dressed by the transformation,~\cite{Mazin:2024p0}
\begin{align}
    \label{eq:dressed_elec}
    \hat{\mathcal{H}}_{\text{e}} \big[ \hbmX \big]
        &= \sum_{pq} h_{pq} \hat{\mathcal{X}}^\dag_p\hat{a}^\dagger_{p} \hat{\mathcal{X}}_q \hat{a}_{q}
        \nonumber \\
        &\ \ + \frac{1}{2} \sum_{pqrs}
            v^{}_{pqrs} \hmX^{\dag}_{p} \hat{a}^\dagger_{p} \hmX^{\dag}_{q} \hat{a}^\dagger_{q} \hmX^{}_{s} \hat{a}_{s} \hmX^{}_{r} \hat{a}_{r}
    \subt{,} \\
\hmX^{\dag}_{p}
    &= \expn{ \frac{f_{\alpha} \cdot \eta^{\alpha}_{p} }
                        {\sqrt{2 {\omega}_{\alpha}}}
                ( \hat{b}^\dagger_{\alpha} - \hat{b}_{\alpha} )
            }
    \subt{,}
\end{align}
where the elements of $\boldsymbol{\eta}^{\alpha}$ are orbital-specific coherent parameters corresponding to the eigenvalues of the dipole-coupling matrix $(\boldsymbol{\lambda}_{\alpha} \cdot \hat{\boldsymbol{\mu}})$. The last two terms of Eq.~\ref{eq:eff_pf_ham} are the residual bilinear-coupling and dipole self-energy terms, proportional to $\Delta\boldsymbol{\lambda}_{\alpha}$ and $\Delta\boldsymbol{\lambda}^{2}_{\alpha}$, respectively. Both terms vanish in the fully displaced limit, $f_{\alpha}=1$.

Effective Hamiltonians for each subsystem can be constructed by performing partial traces over the complementary degrees of freedom. Tracing out the electronic subsystem yields an effective photonic Hamiltonian,
\begin{align}
    \label{eq:eff_ph}
    \tilde{H}_{\text{ph}}
        &= \expval*{\Psi_{\text{e}} | \hmH_{\text{PF}} | \Psi_{\text{e}}}
    \nonumber \\
        &= \hH^{}_{\text{ph}}  + \sqrt{\frac{{\omega}_{\alpha}}{2}}
            ( \Delta\boldsymbol{\lambda}_{\alpha} \cdot \expval*{\hat{\boldsymbol{\mu}}} ) ( \hat{b}^\dagger_{\alpha} + \hat{b}_{\alpha} )
    \nonumber \\
        &\ \ + \frac{1}{2} ( \Delta\boldsymbol{\lambda}_{\alpha} \cdot \expval*{\hat{\boldsymbol{\mu}}} )^{2}
        + \expval*{\Psi_{\text{e}} | \hat{\mathcal{H}}_{\text{e}} \big[ \hbmX \big] | \Psi_{\text{e}}}
        \subt{,}
\end{align}
whereas tracing out the photonic degrees of freedom produces an effective electronic Hamiltonian,
\begin{align}
    \label{eq:eff_elec}
    \tilde{H}_{\text{e}}
        &= \expval*{\Psi_{\text{ph}} | \hmH_{\text{PF}} | \Psi_{\text{ph}}}
    \nonumber \\
        &= \sum_{pq} G^{\alpha}_{pq} h^{}_{pq} \hat{a}^\dagger_{p} \hat{a}_{q}
            + \frac{1}{2} \sum_{pqrs} G^{\alpha}_{pqrs}
            v^{}_{pqrs} \hat{a}^\dagger_{p} \hat{a}^\dagger_{q} \hat{a}_{s} \hat{a}_{r}
    \nonumber \\
        &\ \ + (1 - f_{\alpha}) \sqrt{\frac{{\omega}_{\alpha}}{2}}
            ( \boldsymbol{\lambda}_{\alpha} \cdot \hat{\boldsymbol{\mu}} ) \cdot
            q_\alpha
    \nonumber \\
        &\ \   + \frac{(1-f_{\alpha})^{2}}{2} ( \boldsymbol{\lambda}_{\alpha} \cdot \hat{\boldsymbol{\mu}} {)}^{2}
    \subt{,}
\end{align}
where $q_\alpha\equiv \bra{\Psi_{\text{ph}}}(\hat{b}^\dagger_\alpha + \hat{b}_\alpha)\ket{\Psi_{\text{ph}}}$. The one- and two-electron integrals $h_{pq}$ and $v_{pqrs}$ are dressed by the Gaussian factors $G^{\alpha}_{pq}$ and $G^{\alpha}_{pqrs}$, respectively,
\begin{align}
    \label{eq:g_fact}
    G &=\prod_\alpha e^{\frac{-\mathbf{A}^2_\alpha}{2}}
            \cdot\bra{\Psi^{}_{\text{ph}}} e^{\mathbf{A}^{}_{\alpha} \hat{b}^\dagger_{\alpha}}
            e^{-\mathbf{A}^{}_{\alpha} \hat{b}_{\alpha}} \ket{\Psi^{}_{\text{ph}}}
    \subt{,}
\end{align}
where $\mathbf{A}_{\alpha}$ is a one- or two-body matrix containing differences of the $\boldsymbol{\eta}^{\alpha}$ parameters,
\begin{align}
    \label{eq:pre_factor_A}
    \mathbf{A}^{}_{\alpha}
        &= \frac{-f_{\alpha} \cdot \boldsymbol{\eta}^{\alpha}}
                         {\sqrt{2 {\omega}_{\alpha}}}
    \subt{,} \\
    \label{eq:eta_1e}
    \eta^{\alpha}_{pq}
        &= ( \eta^{\alpha}_{p} - \eta^{\alpha}_{q} )
    \subt{,} \\
    \label{eq:eta_2e}
    \eta^{\alpha}_{pqrs}
        &= ( \eta^{\alpha}_{p} + \eta^{\alpha}_{q}
            - \eta^{\alpha}_{s} - \eta^{\alpha}_{r} )
    \subt{.}
\end{align}

When the photonic subsystem is restricted to the vacuum state, terms in Eq.~\ref{eq:eff_elec} that depend on expectation values of the photonic operators vanish, and the Gaussian factors in Eq.~\ref{eq:g_fact} reduce to the exponential term
\begin{equation}
    \label{eq:exp_term}
    G^{0} = \prod_\alpha e^{\frac{-\mathbf{A}^{2}_{\alpha}}{2}} = e^{\frac{- (f_{\alpha} \cdot \boldsymbol{\eta}^{\alpha} )^{2} }{4 {\omega}_{\alpha}} }.
\end{equation}
A detailed derivation of the Gaussian factors in Eq.~\ref{eq:g_fact} is provided in the Supporting Information.

\subsection{QED Hartree-Fock}
QED Hartree--Fock (QED-HF) extends the traditional HF approach to incorporate interactions between electronic and photonic degrees of freedom. The electronic component $\ket{{\Psi}_{\text{e}}}$ is a standard Hartree--Fock wavefunction: a single Slater determinant that provides a mean-field approximation to the electronic structure,
\begin{align}
    \label{eq:hf_wf}
    \ket{{\Psi}^{}_{\text{e}}}
        &\equiv \ket{{\Psi}^{}_{\text{HF}}}
        = \prod^{N_\text{e}}_{i}
           \hat{a}^\dagger_{i} \ket{{0}^{}_{\text{e}}}
        \subt{.}
\end{align}
Because the effective Hamiltonian in Eq.~\ref{eq:eff_pf_ham} depends on the molecular dipole operator $\hat{\boldsymbol{\mu}}$, it is convenient to introduce the dipole basis,
\begin{align}
    \label{eq:dip_basis}
    \sum_{pq} V^{\dagger}_{px} ( \boldsymbol{\lambda}^{}_{\alpha} \cdot \boldsymbol{\mu}^{}_{pq} ) V^{}_{qy}
        = \sum_{xy} ( \boldsymbol{\lambda}^{}_{\alpha} \cdot \tilde{\boldsymbol{\mu}}^{}_{xy} ) \delta^{}_{xy}
    \subt{,}
\end{align}
where $\boldsymbol{V}$ diagonalizes the matrix $(\boldsymbol{\lambda}_{\alpha}\cdot \boldsymbol{\mu})$. Quantities represented in the dipole basis are denoted with a tilde (e.g., $\tilde{\boldsymbol{\mu}}$, $\tilde{\boldsymbol{h}}$, $\tilde{\boldsymbol{v}}$, and $\tilde{\boldsymbol{\rho}}$).

The total energy functional for the QED-HF methods in the dipole basis is
\begin{align}
    \label{eq:qedhf_en}
   E= &
   \sum_{pq} G^{\alpha}_{pq} \tilde{h}^{}_{pq} \tilde{\rho}^{}_{pq}
    \nonumber \\
        &\ \ + \frac{1}{2} \sum_{pqrs} G^{\alpha}_{pqrs} \tilde{v}^{}_{pqrs}
        ( \tilde{\rho}^{}_{pq} \tilde{\rho}^{}_{rs} - \frac{1}{2} \tilde{\rho}^{}_{ps} \tilde{\rho}^{}_{rq} )
    \nonumber \\
        &\ \ + (1 - f^{}_{\alpha}) \sqrt{\frac{\omega_{\alpha}}{2}} \sum_{pq} 
            ( \boldsymbol{\lambda}^{}_{\alpha} \cdot \boldsymbol{\mu} )^{}_{pq} \rho^{}_{pq}
            \cdot q_{\alpha}
    \nonumber \\
        &\ \ + \frac{(1-f^{}_{\alpha})^{2}}{2} 
            \Big[ \sum_{pq} 
                ( \boldsymbol{\lambda}^{}_{\alpha} \cdot \boldsymbol{\mu} )^{}_{pq}
            \rho^{}_{pq} \Big]^{2} 
    \subt{,}
\end{align}
and is minimized with respect to both the electronic density $\tilde{\boldsymbol{\rho}}$ and the variational parameter $f_{\alpha}$. The expectation value of the sum of photonic creation and annihilation operators is denoted by $q_{\alpha}$. The last two terms in Eq.~\ref{eq:qedhf_en} are residual terms that vanish in the limit $f_{\alpha}\rightarrow 1$.

To characterize cavity-induced electronic changes in the original picture, we evaluate the electronic one-particle density matrix after undoing the variational displacement and tracing out the photons. For a single cavity mode, this gives $\tilde{\rho}_{pq}^{\mathrm{orig}}=\tilde{\rho}_{pq}G_{pq}^{\alpha}$, where $\tilde{\rho}_{pq}=\sum_{\sigma}\langle\Psi_{\text{e}}|\hat{a}_{q\sigma}^{\dagger}\hat{a}_{p\sigma}|\Psi_{\text{e}}\rangle$. Transforming to the atomic-orbital basis gives $\boldsymbol{\rho}^{\mathrm{orig}}=\mathbf{V}\tilde{\boldsymbol{\rho}}^{\mathrm{orig}}\mathbf{V}^{\dagger}$.
At zero coupling, the bare HF density is recovered, although individual natural orbitals remain arbitrary within degenerate occupation subspaces. The orbital-comparison convention used below therefore fixes the zero-coupling gauge with a bare canonical HF orbital and tracks the corresponding finite-coupling natural orbital by overlap.

\section{\label{sec:details}Computational Details}

All QED electronic-structure calculations were performed using the open-source Python quantum chemistry package \textsc{PySCF}~\cite{Sun:2020p024109} together with \textsc{OpenMS}, an in-house code under active development. The \textsc{OpenMS} framework incorporates cavity QED effects into the existing \textsc{PySCF} infrastructure, including integral evaluation, the SCF kernel, and gradient evaluation. Two systems were studied to examine energy-convergence behavior and cavity-coupling effects over a broad range of coupling strengths.

First, calculations were performed on 3-hydroxypropenal, \ce{C3H4O2}, at a fixed geometry (provided in the Supporting Information) using the cc-pVDZ basis set.~\cite{Dunning:1989p1007} A single cavity mode with frequency $\omega = 0.5$ a.u. and a polarization vector along a principal axis of the molecule was employed. The electronic energy and gradient convergence thresholds were set to $10^{-9}$ a.u. and $\sqrt{10^{-9}}$ a.u., respectively, corresponding to the default \textsc{PySCF} thresholds. Three QED self-consistent-field methods were considered: QED-HF, SC-QED-HF, and VT-QED-HF. For each method, the ground-state QED energy was computed as a function of the light--matter coupling strength $\lambda_{\alpha}$, which was varied from $0$ to $0.30$ a.u. The photonic degrees of freedom were represented in a Fock-state basis, and calculations were deemed converged with respect to the number of Fock states, $n$, when the energy difference between successive calculations satisfied $\abs{E^{n+1} - E^{n}} \leq 10^{-6} \text{ a.u.}$ across all coupling strength values. This convergence criterion was used for all methods.

In a separate set of calculations, lithium hydride, \ce{LiH}, was studied at its equilibrium bond length $r_{\ce{Li}-\ce{H}} = 1.6$ \r{A} using the same cc-pVDZ basis set.~\cite{Dunning:1989p1007} The molecule was coupled to a cavity mode with frequency $\omega = 0.5$ a.u. and polarization along the \ce{Li}--\ce{H} bond axis. QED full configuration interaction (QED-FCI) results were used as a reference for the three QED mean-field methods. The same electronic energy and gradient convergence thresholds were applied. For all four \textit{ab initio} methods, the ground-state energy was calculated both with a photonic basis converged according to the criterion above and with a basis restricted to the vacuum state. These calculations were performed for light--matter coupling strengths $\lambda_{\alpha}$ from $0$ to $0.30$ a.u.

\section{\label{sec:results}Results and Discussion}
\subsection{\label{sec:results:c3h4o2}3-Hydroxypropenal}
We analyze the effect of photonic basis size on the energies produced by three QED mean-field methods: QED-HF, SC-QED-HF, and VT-QED-HF.

\begin{figure}[htb]
    \centering
    \includegraphics[width=0.8\linewidth]{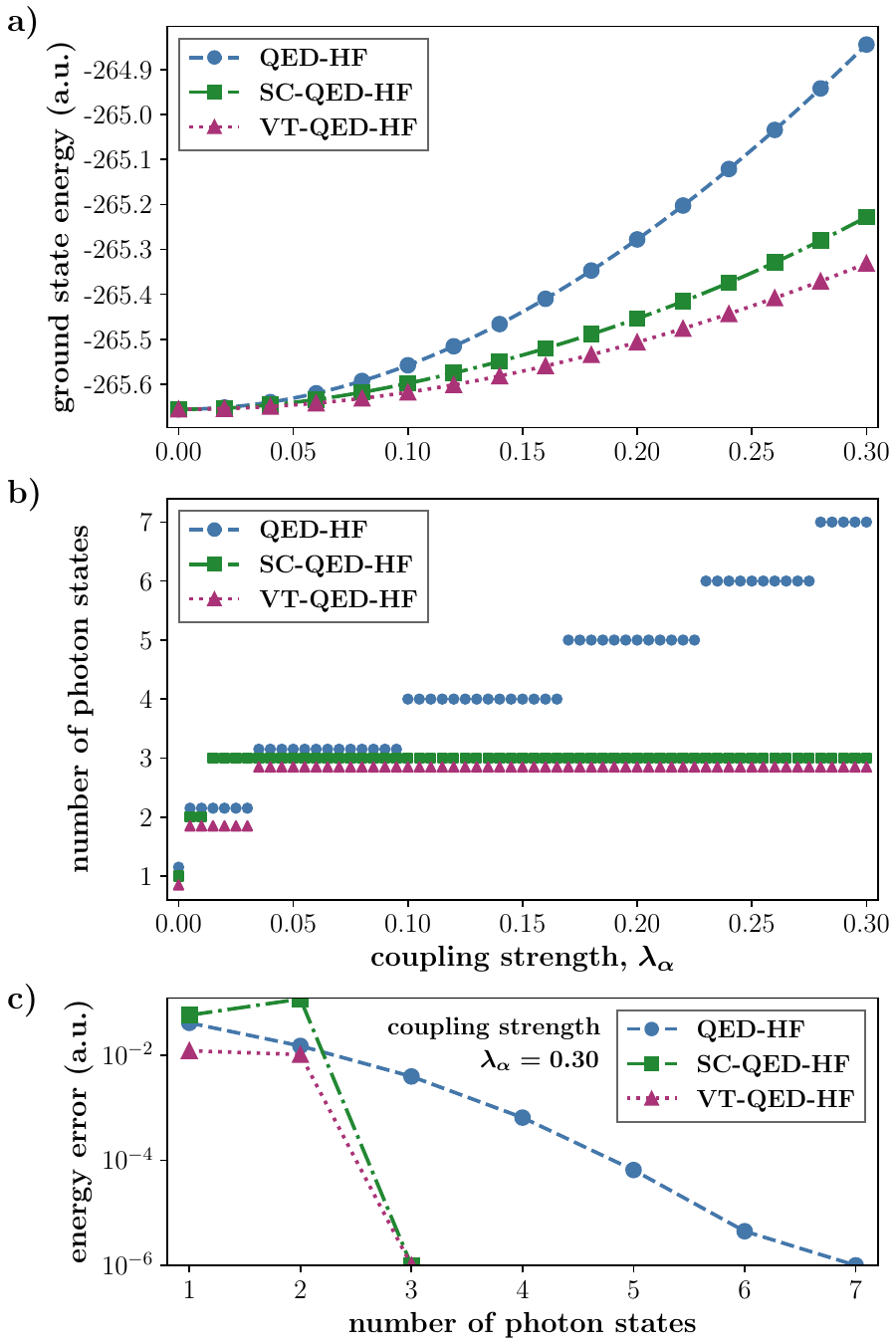}
    \caption{\textbf{(a)} Ground-state energies computed with QED-HF, SC-QED-HF, and VT-QED-HF as functions of the light--matter coupling strength $\lambda_{\alpha}$. \textbf{(b)} Number of photon states required to converge the QED ground-state energies to within $10^{-6}$ a.u. The QED-HF and VT-QED-HF markers (blue circles and red triangles) are shifted vertically relative to the SC-QED-HF markers (green squares) for visual clarity. \textbf{(c)} Ground-state energy convergence of the three QED-HF methods at $\lambda_{\alpha}=0.30$.}
    \label{fig:en_vs_lambda}
\end{figure}

In Fig.~\ref{fig:en_vs_lambda}(a), the ordering of the QED ground-state energy curves is consistent with the variational principle over the entire coupling-strength range. QED-HF produces the highest energies, followed by SC-QED-HF, whereas VT-QED-HF produces the lowest energies at every value of $\lambda_{\alpha}$.

The methods also differ in the number of photonic Fock states required to converge the energy to $10^{-6}$ a.u. [Fig.~\ref{fig:en_vs_lambda}(b)]. At $\lambda_{\alpha}=0.30$, QED-HF requires as many as 7 Fock states to converge the ground-state energy, whereas SC-QED-HF and VT-QED-HF reach the same accuracy with only 3 states. Figure~\ref{fig:en_vs_lambda}(c) shows the corresponding ground-state energy error as a function of the number of Fock states on a $\log_{10}$ scale.

\begin{figure*}[t]
    \centering
    \includegraphics[width=0.90\textwidth]{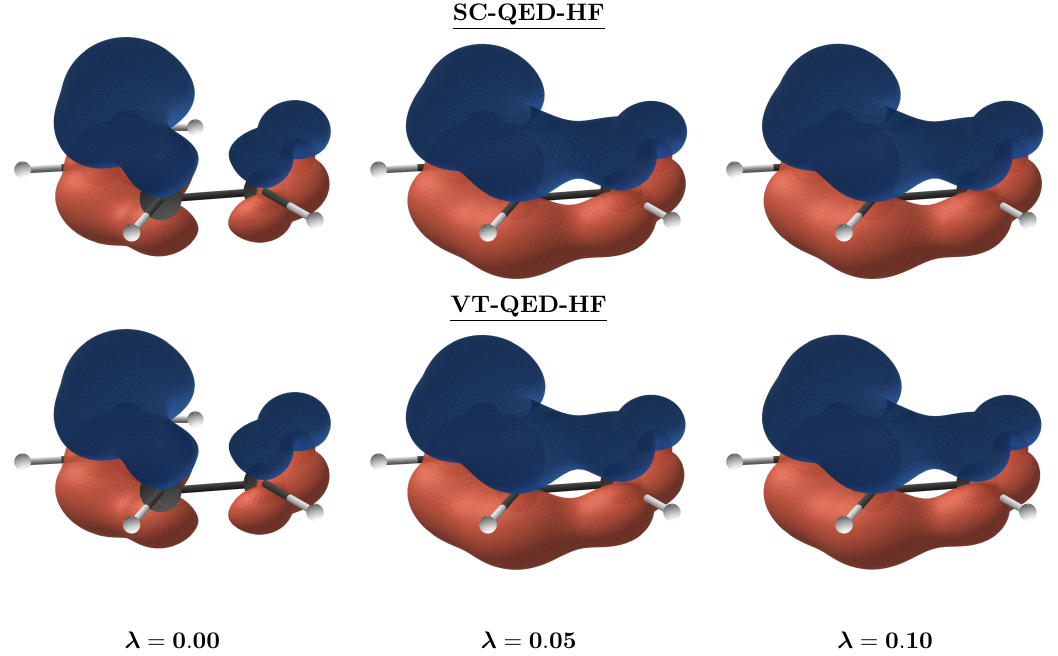}
    \caption{Reference-tracked original-picture natural orbitals of 3-hydroxypropenal (\ce{C3H4O2}) at light--matter coupling strengths $\lambda_{\alpha}=0$, $0.05$, and $0.10$ a.u. The top and bottom rows show SC-QED-HF and VT-QED-HF results, respectively. At zero coupling, the bare-HF canonical MO fixes the gauge within the degenerate occupied natural-orbital subspace. At finite coupling, the natural orbital with the largest AO-metric overlap with this reference is shown. Blue and red denote the positive and negative phases, respectively, at isovalues of $\pm 0.05$ a.u.}
    \label{fig:qed_orbitals}
\end{figure*}

We attribute these energetic differences to the QED-HF methods' distinct ans{\"a}tze. QED-HF employs a product-state ansatz that does not dress the electronic wavefunction with the cavity mode and therefore requires a larger photonic basis when the coherent-state representation is not used. In contrast, the unitary transformations in SC-QED-HF and VT-QED-HF dress the electronic degrees of freedom and incorporate electron--photon correlation into the wavefunction. This variational flexibility permits a more compact photonic basis.

In Fig.~\ref{fig:qed_orbitals}, we compare the reference-tracked $\pi$-bonding natural orbital obtained from the original-picture electronic one-particle density matrix for SC-QED-HF and VT-QED-HF. At $\lambda_{\alpha}=0$, the occupied natural occupations are exactly degenerate, so the bare-HF canonical $\pi$-bonding orbital is used only to fix the orbital gauge. At finite coupling, the plotted natural orbital is selected independently for each method by maximum AO-metric overlap with this reference. No rotations between natural orbitals with distinct occupations are introduced. For SC-QED-HF, the selected occupations are $1.99981$ and $1.99921$ at $\lambda_{\alpha}=0.05$ and $0.10$, with reference overlaps of $0.9725$ and $0.9732$, respectively. The corresponding VT-QED-HF occupations are $1.99996$ and $1.99984$, with reference overlaps of $0.9771$ and $0.9784$.

The selected orbital changes negligibly between $\lambda_{\alpha}=0.05$ and $0.10$: the neighboring-point overlaps are $0.999989$ for SC-QED-HF and $0.999970$ for VT-QED-HF. Moreover, the SC-QED-HF and VT-QED-HF orbitals overlap by $0.999593$ at $\lambda_{\alpha}=0.05$ and $0.999459$ at $0.10$. The two transformed methods therefore predict essentially the same finite-coupling $\pi$-orbital shape over this range. The visible difference between the zero- and finite-coupling panels does not, by itself, imply an abrupt redistribution of the total electronic density. At zero coupling, an individual natural orbital is nonunique within the degenerate occupied subspace, whereas finite coupling lifts that degeneracy and selects a particular natural-orbital orientation. Density redistribution should therefore be assessed using invariant quantities such as the one-particle density and its real-space difference. The orbital plots instead represent the tracked $\pi$ character.

\begin{figure}[htb]
    \centering
    \includegraphics[width=0.9\linewidth]{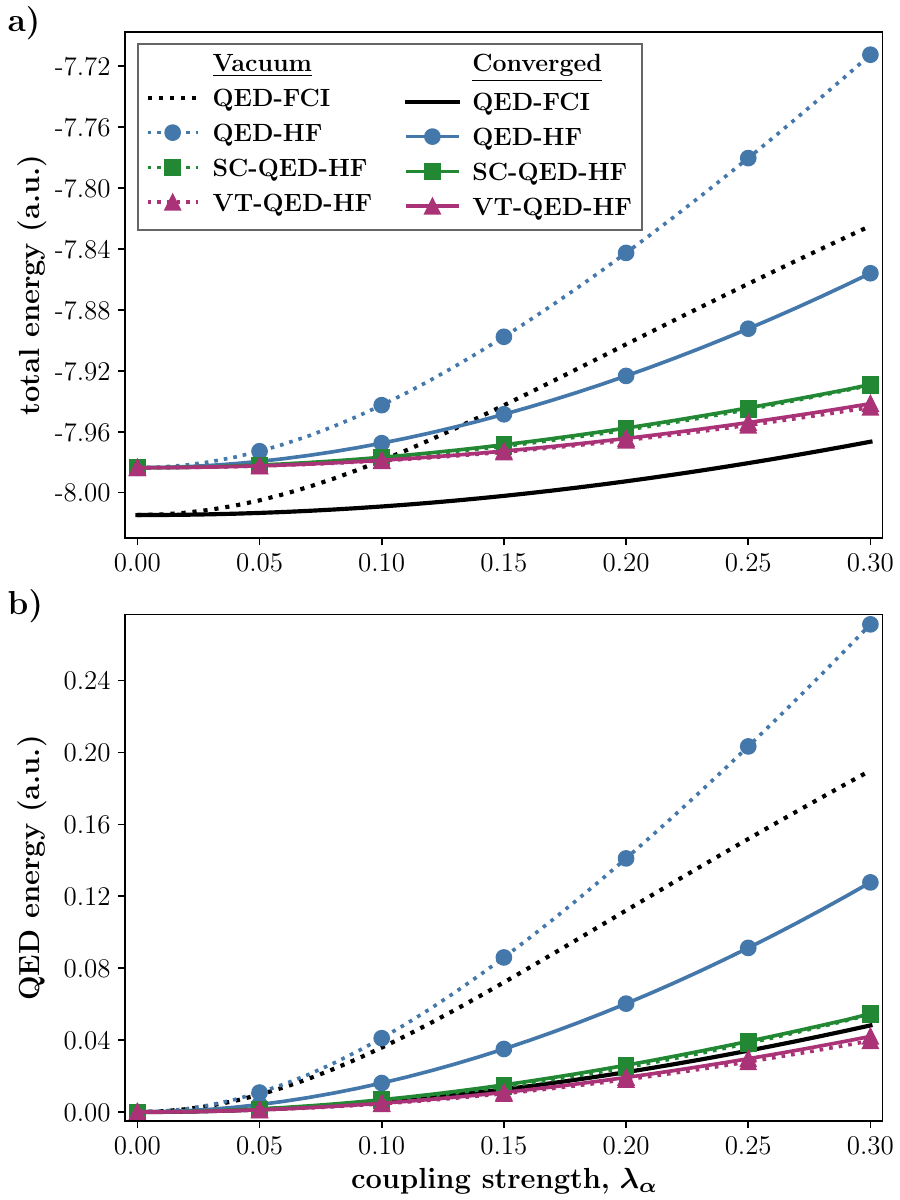}
    \caption{Total and QED energies (defined relative to the $\lambda_{\alpha}=0$ energy) as functions of the light--matter coupling strength $\lambda_{\alpha}$. \textbf{(a)} Ground-state energies computed using only the photonic vacuum state (dashed lines) and converged with respect to the number of Fock states (solid lines). \textbf{(b)} Corresponding QED energy contributions.}
    \label{fig:qed_energy}
\end{figure}

\subsection{\label{sec:results:lih}Lithium Hydride}
Having established the efficacy of unitary transformations in QED mean-field methods, we next examine how photonic-basis truncation affects the electron--photon and dipole self-energy contributions to the total ground-state energy. We consider a lithium hydride molecule coupled to an optical cavity with frequency $\omega = 0.5$ a.u. and polarization along the \ce{Li}--\ce{H} bond axis.

Figure~\ref{fig:qed_energy} illustrates the effect of photonic-basis truncation on the light--matter contribution to the total energy. For each method, we define the QED energy $E_{\text{QED}}$ as the energy relative to the Hartree--Fock energy $E_{\text{HF}}$ at $\lambda_{\alpha}=0$:
\begin{align*}
    E_{\text{QED}}(\lambda_{\alpha}) = E(\lambda_{\alpha}) - E_{\text{HF}}
    \subt{.}
\end{align*}
Calculations were performed both with the photonic basis restricted to the vacuum state $\ket*{0^{\alpha}_{\text{ph}}}$ (dashed lines) and with the basis converged in size (solid lines). The vacuum-only approximation causes QED-FCI and QED-HF to overestimate the total energy as the light--matter coupling strength increases relative to SC-QED-HF and VT-QED-HF, which are less sensitive to photonic-basis truncation. Each solid-line energy is converged to within $10^{-6}$ a.u. with respect to the photonic basis size. As shown in Fig.~\ref{fig:qed_energy}(b), the entangled ans{\"a}tze predict similar QED energy contributions ($\approx 0.04$ a.u.) in the vacuum-only and converged calculations, and their potential energy curves (PECs) are comparatively insensitive to the photonic basis size. By contrast, the QED-HF and QED-FCI results depend more strongly on the basis size: the magnitude of their QED energy contributions decreases by approximately a factor of two upon convergence. The converged QED-FCI PEC becomes more consistent with the SC-QED-HF and VT-QED-HF curves, suggesting that an insufficient photonic basis may cause QED-FCI to exhibit qualitatively mean-field-like behavior. These comparisons emphasize the importance of photonic-basis convergence for capturing qualitative trends in strong light--matter coupling regimes.

\section{\label{sec:summary}Summary}
In this work, we developed a generalized QED self-consistent-field (QED-SCF) framework for strongly coupled light--matter systems that extends QED Hartree--Fock theory to an arbitrary number of photonic Fock states. The framework provides a common formulation of standard QED-HF and two variational approaches: strongly coupled QED-HF (SC-QED-HF) and variationally transformed QED-HF (VT-QED-HF). Using this framework, we characterized convergence with respect to the photonic basis and showed that the methods have substantially different basis requirements because they represent electron--photon correlation differently.

The results in Sec.~\ref{sec:results:c3h4o2} demonstrate that the unitary transformations coupling the electronic and photonic degrees of freedom in SC-QED-HF and VT-QED-HF provide a substantial advantage over the direct-product QED-HF ansatz. Whereas QED-HF requires up to 7 photonic Fock states to achieve $10^{-6}$ a.u. convergence at $\lambda_{\alpha}=0.30$, the transformed methods require only 3 states to reach comparable accuracy. We attribute this efficiency to the explicit dressing of the electronic wavefunction by the photonic degrees of freedom, which incorporates electron--photon correlation into the variational ansatz.

The lithium hydride calculations in Sec.~\ref{sec:results:lih} further show the physical implications of photonic-basis truncation. When restricted to the vacuum state, QED-HF and QED-FCI substantially overestimate the QED energy contribution relative to SC-QED-HF and VT-QED-HF. Upon convergence with respect to the photonic basis size, the QED energy contributions predicted by QED-HF and QED-FCI decrease by approximately a factor of two, and the QED-FCI potential energy curve becomes qualitatively consistent with those of the two transformed mean-field methods. This behavior suggests that an insufficient photonic basis may cause QED-FCI to exhibit mean-field-like behavior and obscure important electron--photon correlation effects.

Our results highlight the importance of photonic-basis convergence for obtaining qualitatively correct cavity QED simulations. The additional variational flexibility introduced through the displacement operators in SC-QED-HF and VT-QED-HF enables efficient treatment of electron--photon correlation in the strong-coupling regime. The compact photonic bases required by these methods reduce computational cost and improve energetic predictions when the photonic basis is incomplete.

\section{\label{sec:ack}Acknowledgments}
\begin{acknowledgments}
We acknowledge support from the US DOE, Office of Science, Basic Energy Sciences, Chemical Sciences, Geosciences, and Biosciences Division under Triad National Security, LLC (``Triad") contract Grant 89233218CNA000001 (FWP: LANLECF7). This research used computational resources provided by the Institutional Computing (IC) Program and the Darwin testbed at Los Alamos National Laboratory (LANL), which is funded by the Computational Systems and Software Environments subprogram of LANL's Advanced Simulation and Computing program. LANL is operated by Triad National Security, LLC, for the National Nuclear Security Administration of the U.S. Department of Energy (Contract No. 89233218CNA000001). 
\end{acknowledgments}

{\bf Author contributions:}

{\bf Data availability:}
The data supporting this study's findings are available from the corresponding author upon request.

{\bf Code availability:}
The developed code used for this study is implemented in the \textsc{OpenMS} program, which is available at ``\url{https://github.com/lanl/OpenMS}". Sample inputs for the calculations performed in this manuscript are available from the corresponding author upon request.

\bibliography{qednphoton}
\appendix
\begin{widetext}
\section{\label{sec:app1}Expectation Value and gradient of the Displacement Operator}
\subsection{Expectation value}
The calculation of the Gaussian factors in main text that scale the electronic degrees of freedom requires computing the expectation value of the photonic operators introduced by the displacement operator $\hat{D} ( \hbmu, f_{\alpha} )$ used to perform the unitary transformation in Eq.~13 of main text, 
\begin{equation}
    \label{eq:app_gfact}
        G = \expval*{\Psi^{}_{\text{ph}} \lvert
            \expn{ \frac{-f_{\alpha} \cdot ( \boldsymbol{\lambda}_{\alpha} \cdot \hat{\boldsymbol{\mu}} )}
                         {\sqrt{2 {\omega}_{\alpha}}}   
            ( \hat{b}^\dagger_{\alpha} - \hat{b}_{\alpha} )} \rvert \Psi^{}_{\text{ph}}}
\end{equation}
Using the Baker-Campbell-Hausdorff formula, equation Eq.~\ref{eq:app_gfact} can be expressed as:
\begin{align}
    \label{eq:app_bch}
    G
    &= \expval*{\Psi^{}_{\text{ph}} \lvert \expn{\mathbf{A}^{}_{\alpha} ( \hat{b}^\dagger_{\alpha} - \hat{b}_{\alpha} )} \rvert \Psi^{}_{\text{ph}}}
    \nonumber \\
    &= \expn{\frac{-\mathbf{A}^{2}_{\alpha}}{2}}
        \cdot \expval*{\Psi^{}_{\text{ph}} \lvert \expn{\mathbf{A}^{}_{\alpha} \hat{b}^\dagger_{\alpha}}
        \cdot \expn{-\mathbf{A}^{}_{\alpha} \hat{b}_{\alpha}} \rvert \Psi^{}_{\text{ph}}}
    \subt{,}
\end{align}
in which the $\mathbf{A}_{\alpha}$ terms are the pre-factors proportional to dipole-interaction matrix $(\boldsymbol{\lambda}_{\alpha} \cdot \hbmu)$.

The expectation value of the scaled photonic creation and annihilation operators ($e^{\mathbf{A}^{}_{\alpha} \hat{b}^\dagger_{\alpha}}$ and $e^{-\mathbf{A}^{}_{\alpha} \hat{b}_{\alpha}}$, respectively) is computed with respect to the photonic degrees of freedom. If the photonic basis $\ket*{\Psi_{\text{ph}}}$ includes any Fock states $\ket*{n}$ alongside the photon vacuum state $\ket*{0^{\alpha}_{\text{ph}}}$, the expectation value of the photonic operators in the second term of Eq.~\ref{eq:app_bch} is non-zero. The action of $e^{-\mathbf{A} \hat{b}_{\alpha}}$ and $e^{\mathbf{A} \hat{b}^\dagger_{\alpha}}$ on the Fock states generate the Laguerre polynomials:
\begin{align}
    \label{eq:app_boson_des}
    e^{ -\mathbf{A}^{}_{\alpha} \hat{b}_{\alpha} } \ket{n}
        &= \sum^{n}_{k} \frac{( -\mathbf{A}^{}_{\alpha} )^{k}}{k!}
            \cdot \sqrt{ \frac{n!}{(n - k)!} } \ \ket{n - k}
    \subt{,} \\
    \label{eq:app_boson_cre}
    e^{ \mathbf{A}^{}_{\alpha} \hat{b}^\dagger_{\alpha} } \ket{n - k}
        &= \sum^{\infty}_{l} \frac{ ( \mathbf{A}^{}_{\alpha} )^{l} }{l}
            \cdot \sqrt{ \frac{(n - k + l)!}{(n - k)!} } \ \ket{n - k + l}
    \subt{.}
\end{align}
The associated Laguerre polynomials $\mathcal{L}^{(m-n)}_{n} \left[ \mathbf{x} \right]$ are defined as
\begin{align}
    \label{eq:app_laguerre}
    \mathcal{L}^{j}_{n} \left[ \mathbf{x} \right]
        = \sum_{k=0}^{n} \frac{(n + j)!}{(n - k)!\cdot (j + k)! \cdot k!}
        \cdot ( -\mathbf{x} {)}^{k}
    \subt{.}
\end{align}
With $\mathbf{x} \equiv \lvert\lvert \mathbf{A}_{\alpha} \rvert\rvert^{2}$, Eq.~\ref{eq:app_laguerre} is used to write down the matrix elements of the final expectation value, where $m = n - k + l$ (therefore, $l = m - n + k$):
\begin{align}
    \label{eq:app_vtqed_me}
    \mathbf{G}
        &= \expval*{\Psi^{}_{\text{ph}} \lvert \expn{\mathbf{A}^{}_{\alpha}
                    ( \hat{b}^\dagger_{\alpha} - \hat{b}_{\alpha} )} \rvert \Psi^{}_{\text{ph}}}
    \nonumber \\
        &= \expn{\frac{-\mathbf{A}^{2}_{\alpha}}{2}}
            \ \ \cdot \sum^{\min(m,n)}_{k}
            \sqrt{ \frac{n!}{(n - k)!} \cdot \frac{m!}{(n - k)!} }
            \cdot \frac{( \mathbf{A}^{}_{\alpha} )^{m-n+k} \cdot (-\mathbf{A}^{}_{\alpha})^{k}}
                        {k! \cdot (m - n + k)!}
    \nonumber \\
        &= \expn{\frac{-\mathbf{A}^{2}_{\alpha}}{2}}
            \cdot \sum^{m \geq n}_{mn}\sqrt{\frac{n!}{m!}} \cdot \mathbf{A}^{(m - n)}_{\alpha}
            \cdot \mathcal{L}^{(m-n)}_{n}
            \Big[ \lvert\lvert \mathbf{A}^{}_{\alpha} \rvert\rvert^{2} \Big]
\end{align}

\section{\label{sec:app2}Gradient}
The gradient expressions include additional terms to account for the energy changes with respect to the $\boldsymbol{\eta}^{\alpha}$ and $f_{\alpha}$ parameters, collectively denoted as $\mathbf{X}$:
\begin{align}
    \label{eq:app_grad}
    \frac{\partial E}{\partial \mathbf{X}}
        &= \frac{\partial}{\partial \mathbf{X}} \Biggr(
            \expn{\frac{-\mathbf{A}^{2}_{\alpha}}{2}}
            \cdot \sqrt{\frac{n!}{m!}} \cdot \mathbf{A}^{(m - n)}_{\alpha}
            \cdot \mathcal{L}^{(m-n)}_{n}
            \Big[ \lvert\lvert \mathbf{A}^{}_{\alpha} \rvert\rvert^{2} \Big]
           \Biggr)
        \subt{,}
\end{align}
since $\mathbf{A}_{\alpha} \equiv \mathbf{A}_{\alpha} ( \boldsymbol{\eta}^{\alpha}, f_{\alpha})$. Expanding Eq.~\ref{eq:app_grad},
\begin{align}
    \frac{\partial E}{\partial \mathbf{X}}
        &= \frac{\partial}{\partial \mathbf{X}}
            \Biggr( \expn{\frac{-\mathbf{A}^{2}_{\alpha}}{2}} \Biggr) \cdot
            \Biggr(
                \sqrt{\frac{n!}{m!}} \cdot \mathbf{A}^{(m - n)}_{\alpha}
                \cdot \mathcal{L}^{(m-n)}_{n}
                \Big[ \lvert\lvert \mathbf{A}^{}_{\alpha} \rvert\rvert^{2} \Big]
            \Biggr)
    \nonumber \\
    \label{eq:app_grad_term}
        &\ \ + \Biggr( \expn{\frac{-\mathbf{A}^{2}_{\alpha}}{2}} \Biggr)
            \cdot \frac{\partial}{\partial \mathbf{X}}
            \Biggr(
                \sqrt{\frac{n!}{m!}} \cdot \mathbf{A}^{(m - n)}_{\alpha}
                \cdot \mathcal{L}^{(m-n)}_{n}
                \Big[ \lvert\lvert \mathbf{A}^{}_{\alpha} \rvert\rvert^{2} \Big]
            \Biggr)
    \subt{,}
\end{align}
produces two derivative terms. The derivative in the second term of Eq.~\ref{eq:app_grad_term} evaluates to
\begin{align}
    \label{eq:app_grad_term2}
    &= \frac{\partial}{\partial \mathbf{X}}
        \Biggr(
            \sqrt{\frac{n!}{m!}} \cdot \mathbf{A}^{(m - n)}_{\alpha}
            \cdot \mathcal{L}^{(m-n)}_{n}
            \Big[ \lvert\lvert \mathbf{A}^{}_{\alpha} \rvert\rvert^{2} \Big]
        \Biggr)
    \nonumber \\
    &= \sqrt{\frac{n!}{m!}} \cdot
        \Biggr[
            \frac{\partial}{\partial \mathbf{X}}
            \Biggr(
                \mathbf{A}^{(m - n)}_{\alpha}
            \Biggr)
            \cdot \mathcal{L}^{(m-n)}_{n}
            \Big[ \lvert\lvert \mathbf{A}^{}_{\alpha} \rvert\rvert^{2} \Big]
            + \mathbf{A}^{(m - n)}_{\alpha}
            \cdot \frac{\partial}{\partial \mathbf{X}}
            \Biggr(
                \mathcal{L}^{(m-n)}_{n}
                \Big[ \lvert\lvert \mathbf{A}^{}_{\alpha} \rvert\rvert^{2} \Big]
            \Biggr)
        \Biggr]
\end{align}
and requires taking a derivative of the Laguerre polynomial $\mathcal{L}^{(m-n)}_{n} \Big[ \lvert\lvert \mathbf{A}_{\alpha} \rvert\rvert^{2} \Big]$,
\begin{align}
    \label{eq:app_laguerre_deriv}
    \frac{\partial}{\partial \mathbf{X}}
        \Biggr(
            \mathcal{L}^{(m-n)}_{n}
            \Big[ \lvert\lvert \mathbf{A}^{}_{\alpha} \rvert\rvert^{2} \Big]
        \Biggr)
        &= - \Biggr(
                \mathcal{L}^{(m-n+1)}_{n-1}
                \Big[ \lvert\lvert \mathbf{A}^{}_{\alpha} \rvert\rvert^{2} \Big]
            \Biggr)
            \cdot \frac{\partial}{\partial \mathbf{X}}
            \Biggr(
                \Big[ \lvert\lvert \mathbf{A}^{}_{\alpha} \rvert\rvert^{2} \Big]
            \Biggr)
    \subt{,} \\
    \lvert\lvert \mathbf{A}^{}_{\alpha} \rvert\rvert^{2}
        &= \frac{( f_{\alpha} \cdot \boldsymbol{\lambda}_{\alpha} \cdot \boldsymbol{\eta}^{\alpha} {)}^{2}}
                 {2 \omega}
\end{align}

The final expression for the additional energy gradient terms is:
\begin{align}
    \label{eq:app_final_grad}
    &= \frac{\partial}{\partial \mathbf{X}}
        \Biggr( \expn{\frac{-\mathbf{A}^{2}_{\alpha}}{2}} \Biggr) \cdot
        \Biggr(
            \sqrt{\frac{n!}{m!}} \cdot \mathbf{A}^{(m - n)}_{\alpha}
            \cdot \mathcal{L}^{(m-n)}_{n}
            \Big[ \lvert\lvert \mathbf{A}^{}_{\alpha} \rvert\rvert^{2} \Big]
        \Biggr)
    \nonumber \\
    &\ \ + \Biggr( \expn{\frac{-\mathbf{A}^{2}_{\alpha}}{2}} \Biggr)
        \cdot
        \Biggr(
            \sqrt{\frac{n!}{m!}}
            \cdot \frac{\partial}{\partial \mathbf{X}}
            \Biggr(
                \mathbf{A}^{(m - n)}_{\alpha}
            \Biggr)
            \cdot \mathcal{L}^{(m-n)}_{n}
            \Big[ \lvert\lvert \mathbf{A}^{}_{\alpha} \rvert\rvert^{2} \Big]
        \Biggr)
    \nonumber \\
    &\ \ + \Biggr( \expn{\frac{-\mathbf{A}^{2}_{\alpha}}{2}} \Biggr)
        \cdot
        \Biggr(
            \sqrt{\frac{n!}{m!}} \cdot \mathbf{A}^{(m - n)}_{\alpha}
            \cdot \frac{\partial}{\partial \mathbf{X}}
            \Biggr(
                \mathcal{L}^{(m-n)}_{n}
                \Big[ \lvert\lvert \mathbf{A}^{}_{\alpha} \rvert\rvert^{2} \Big]
            \Biggr)
        \Biggr)
    \subt{.}
\end{align}
For systems approximating the photonic subsystem with only the photon vacuum state, the second and third terms of Eq.~\ref{eq:app_final_grad} will evaluate to zero. In other words, while the first term will always impact the energy gradient, the next two terms only contribute to the gradient if the photonic basis set includes Fock states beyond the photon vacuum.

Next, we define the Fock matrix elements in the dipole basis,
\begin{align}
    \label{eq:qedhf_fock}
    &\tilde{F}^{}_{pq} = \frac{\partial E[ \tilde{\boldsymbol{\rho}}, f_{\alpha}]}{\partial \tilde{\rho}^{}_{pq}} 
        = \sum_{pq} G^{\alpha}_{pq} \tilde{h}^{}_{pq}
    + \frac{1}{2} \sum_{rs} G^{\alpha}_{pqrs}
        ( \tilde{v}^{}_{pqrs} - \frac{1}{2} \tilde{v}^{}_{psrq} ) \tilde{\rho}^{}_{rs}
    \nonumber \\
        &\ \ + f^{}_{\alpha} \sqrt{\frac{\omega_{\alpha}}{2}}
            \Big[ 
                ( \boldsymbol{\lambda}^{}_{\alpha} \cdot \tilde{\boldsymbol{\mu}} )^{}_{pp}
                - \eta^{}_{p}
            \Big] \delta^{}_{pq} \cdot q_{\alpha}
            + \frac{f^{2}_{\alpha}}{2}
            \Big[ 
                ( \boldsymbol{\lambda}^{}_{\alpha} \cdot \tilde{\boldsymbol{\mu}} )^{}_{pp}
                - \eta^{}_{p}
            \Big]^{2} \delta^{}_{pq}
    \nonumber \\
        &\ \ + f^{2}_{\alpha}
            \Big[ 
                ( \boldsymbol{\lambda}^{}_{\alpha} \cdot \tilde{\boldsymbol{\mu}} )^{}_{pp}
                - \eta^{}_{p}
            \Big] \delta^{}_{pq}
            \sum_{r} \tilde{\rho}^{}_{rr}
            \Big[ 
                ( \boldsymbol{\lambda}^{}_{\alpha} \cdot \tilde{\boldsymbol{\mu}} )^{}_{r}
                - \eta^{}_{r}
            \Big]
     - \frac{f^{2}_{\alpha}}{2}
            \Big[ 
                ( \boldsymbol{\lambda}^{}_{\alpha} \cdot \tilde{\boldsymbol{\mu}} )^{}_{pp}
                - \eta^{}_{p}
            \Big]
            \Big[ 
                ( \boldsymbol{\lambda}^{}_{\alpha} \cdot \tilde{\boldsymbol{\mu}} )^{}_{qq}
                - \eta^{}_{q}
            \Big] \tilde{\rho}^{}_{pq}
    \nonumber \\
        &\ \ + (1 - f^{}_{\alpha}) \sqrt{\frac{\omega_{\alpha}}{2}} \sum_{pq} 
            ( \boldsymbol{\lambda}^{}_{\alpha} \cdot \boldsymbol{\mu} )^{}_{pq} \delta^{}_{pq}
            \cdot q_{\alpha}
     + \frac{(1-f^{}_{\alpha})^{2}}{2} 
            \Big[ \sum_{pq} 
                ( \boldsymbol{\lambda}^{}_{\alpha} \cdot \boldsymbol{\mu} )^{}_{pq}
            \delta^{}_{pq} \Big]^{2} 
    \subt{.}
\end{align}

In order to variationally minimize the electron-photon coupling, additional energy derivative terms with respect to coherent parameters $\boldsymbol{\eta}^{\alpha}$ and the variational parameter $f_{\alpha}$. Although $\boldsymbol{\eta}^{\alpha}$ are not variational parameters, they are a function of the molecular dipole operator,
\begin{align}
    \label{eq:eta_dipole}
    \sum_{pq} ( \boldsymbol{\lambda}^{}_{\alpha} \cdot \boldsymbol{\mu} )^{}_{pq} \rho^{}_{pq}
        = \sum_{x} \eta^{\alpha}_{x} \tilde{\rho}^{}_{xx}
    \subt{,}
\end{align}
and change nonlinearly. 
The gradient with the respect to $\boldsymbol{\eta}^{\alpha}$ is
\begin{align}
    \label{eq:eta_grad}
    \frac{\partial E[ \tilde{\boldsymbol{\rho}}, f_{\alpha}]}{\partial \eta^{\alpha}_{p}} 
        = & \sum_{q} \frac{\partial G^{\alpha}_{pq}}{\partial \eta^{\alpha}_{p}}
        \tilde{h}^{}_{pq} \tilde{\rho}^{}_{pq}
     + \frac{1}{2} \sum_{qrs}
            \frac{\partial G^{\alpha}_{pqrs}}{\partial \eta^{\alpha}_{p}}
            \tilde{v}^{}_{pqrs} 
            ( \tilde{\rho}^{}_{pq} \tilde{\rho}^{}_{rs}
                - \frac{1}{2} \tilde{\rho}^{}_{ps} \tilde{\rho}^{}_{rq} )
    \nonumber \\
        & - f^{2}_{\alpha}
            \Big[ 
                ( \boldsymbol{\lambda}^{}_{\alpha} \cdot \tilde{\boldsymbol{\mu}} )^{}_{pp}
                - \eta^{}_{p}
            \Big] \tilde{\rho}^{}_{pp}
        - f^{2}_{\alpha} \sum_{q}
            \Big[ 
                ( \boldsymbol{\lambda}^{}_{\alpha} \cdot \tilde{\boldsymbol{\mu}} )^{}_{qq}
                - \eta^{}_{q}
            \Big]
            ( \tilde{\rho}^{}_{pp} \tilde{\rho}^{}_{qq}
                - \frac{1}{2} \tilde{\rho}^{2}_{pq} )
    \subt{,}
\end{align}
while the energy gradient with respect to the parameter $f_{\alpha}$ is
\begin{align}
    \label{eq:fopt_grad}
    \frac{\partial E[ \tilde{\boldsymbol{\rho}}, f_{\alpha}]}{\partial f^{}_{\alpha}} 
        = & \sum_{pq} \frac{\partial G^{\alpha}_{pq}}{\partial f^{}_{\alpha}}
        \tilde{h}^{}_{pq} \tilde{\rho}^{}_{pq}
     + \frac{1}{2} \sum_{pqrs}
            \frac{\partial G^{\alpha}_{pqrs}}{\partial \eta^{\alpha}_{p}}
            \tilde{v}^{}_{pqrs} 
            ( \tilde{\rho}^{}_{pq} \tilde{\rho}^{}_{rs}
                - \frac{1}{2} \tilde{\rho}^{}_{ps} \tilde{\rho}^{}_{rq} )
                + f^{}_{\alpha} \sum_{p}
            \Big[ 
                ( \boldsymbol{\lambda}^{}_{\alpha} \cdot \tilde{\boldsymbol{\mu}} )^{}_{pp}
                - \eta^{}_{p}
            \Big]^{2} \tilde{\rho}^{}_{pp}
    \nonumber\\ &
     + f^{}_{\alpha} \Biggr( \sum_{pq}
            \Big[ 
                ( \boldsymbol{\lambda}^{}_{\alpha} \cdot \tilde{\boldsymbol{\mu}} )^{}_{pp}
                - \eta^{}_{p}
            \Big]
            \Big[ 
                ( \boldsymbol{\lambda}^{}_{\alpha} \cdot \tilde{\boldsymbol{\mu}} )^{}_{qq}
                - \eta^{}_{q}
            \Big]
            \cdot ( \tilde{\rho}^{}_{pp} \tilde{\rho}^{}_{qq}
                - \frac{1}{2} \tilde{\rho}^{2}_{pq} ) \Biggr)
     - (1-f^{}_{\alpha}) 
            \Big[ \sum_{pq} 
                ( \boldsymbol{\lambda}^{}_{\alpha} \cdot \boldsymbol{\mu} )^{}_{pq}
            \rho^{}_{pq} \Big]^{2} 
    \subt{,}
\end{align}
Detailed derivations for the Gaussian factor derivatives $\partial \boldsymbol{G}^{\alpha}/\partial \eta^{\alpha}_{p}$ and $\partial \boldsymbol{G}^{\alpha}/\partial f^{}_{\alpha}$ can be found in the Appendix~\ref{sec:app1}. 

\end{widetext}

\end{document}